\documentclass[showpacs, amsmath, amssymb, amsthm, aps, jmp, thmsb, twocolumn, superscriptaddress]{revtex4-1} 

\usepackage{color}
\usepackage{float}
\usepackage{amsmath}
\usepackage{amssymb}
\usepackage{amsfonts}
\usepackage{amsbsy}
\usepackage{latexsym}
\usepackage{mathrsfs}
\usepackage{bbm}
\usepackage{bm}
\usepackage[dvipdfmx]{graphicx}

\newcommand{\be}{\begin{eqnarray}}
\newcommand{\ee}{\end{eqnarray}}
\def\({\left(}
\def\){\right)}
\def\[{\left[}
\def\]{\right]}

\def\R{\mathbb{R}}

\newcommand{\e}{\mathrm{e}}
\newcommand{\im}{\mathrm{i}}

\newcommand{\dd}{\mathrm{d}}

\newcommand{\bra}[1]{\langle #1 |}
\newcommand{\ket}[1]{| #1 \rangle}

\newcommand{\Tr}{\mathrm{Tr}}
\newcommand{\sla}[1]{\rlap{\kern .15em /}#1}

\begin{document}
\title{Designing Robust Unitary Gates: Application to Concatenated Composite Pulse}

\author{Tsubasa Ichikawa}
\affiliation{Research Center for Quantum Computing, Interdisciplinary
Graduate School of Science and Engineering, Kinki University, 3-4-1
Kowakae, Higashi-Osaka, Osaka 577-8502, Japan}
\author{Masamitsu Bando}
\affiliation{Research Center for Quantum Computing, Interdisciplinary
Graduate School of Science and Engineering, Kinki University, 3-4-1
Kowakae, Higashi-Osaka, Osaka 577-8502, Japan}
\author{Yasushi Kondo}
\affiliation{Research Center for Quantum Computing, Interdisciplinary
Graduate School of Science and Engineering, Kinki University, 3-4-1
Kowakae, Higashi-Osaka, Osaka 577-8502, Japan}
\affiliation{Department of Physics, Kinki University, 3-4-1 Kowakae,
Higashi-Osaka, Osaka 577-8502, Japan}
\author{Mikio Nakahara}
\affiliation{Research Center for Quantum Computing, Interdisciplinary
Graduate School of Science and Engineering, Kinki University, 3-4-1
Kowakae, Higashi-Osaka, Osaka 577-8502, Japan}
\affiliation{Department of Physics, Kinki University, 3-4-1 Kowakae,
Higashi-Osaka, Osaka 577-8502, Japan}

\begin{abstract}
We propose a simple formalism to design unitary gates robust 
against given systematic errors. This formalism generalizes our previous 
observation [Y.~Kondo and M.~Bando, J. Phys. Soc. Jpn. {\bf 80}, 054002
(2011)] that vanishing dynamical phase in some
composite gates is essential to suppress pulse-length errors.
By employing our formalism, we derive a new composite 
unitary gate which can be seen as a concatenation of two known 
composite unitary operations. The obtained unitary gate has 
high fidelity over a wider range of error strengths compared to
existing composite gates.
\end{abstract}
\pacs{03.65.Vf, 03.67.Pp, 82.56.Jn.}

\maketitle

\section{Introduction}

Noise and errors are obstacles against reliable control of a quantum
system. Noise, i.e., random unwanted disturbance to a quantum system we
concern, has been attracting much attention of theoreticians. Many
ideas to suppress noise have been proposed \cite{NC00, NO08} 
in quantum computation, which requires precise control of 
quantum systems \cite{Gaitan07}.
Geometric quantum gates (GQGs) \cite{Zanardi98, Zhu02, Blais03,
Zhu05, Ota09b, Kondo10, Zhu03, Solinas03, Ota08}, that are based on
holonomy \cite{Berry84, Wilczek84, Aharonov87, Page87, Anandan88, Shapere89,gtp,
Mead92, Ota09a}, are such examples. On the other hand, errors, i.e. 
systematic imperfection in control parameters, have also been attracting attention, 
due to their importance in realistic situations. 

To tackle the latter problem, one may decompose a given unitary gate
into a sequence of several unitary operations, whose 
time-ordered product reproduces
the given unitary gate \cite{Levitt81, Counsell85, Tycko85, SP87, 
Levitt96, Freeman99, Claridge99, Cummins00,
Cummins03, Mottonen06, Alway07}. Then the 
sequence becomes robust against given systematic errors by tuning the 
parameters in the constituent unitary operations. 
Such sequences for a two-level system is well-known as composite pulses in NMR
and have been designed by employing various 
techniques, such as the Magnus expansion \cite{SP87} and 
quaternion algebra \cite{Cummins03}, for example.
In the following, we often use a ``composite gate'' to denote a
composite pulse when it is regarded as a quantum gate.

As mentioned in \cite{Counsell85}, there are several lines of thought to
understand composite pulses in a unified manner. Motivated by these, 
we have proved in \cite{Kondo10} that GQGs for a two-level system are 
insensitive to an error in the amplitude of the control parameters and 
shown that many existing composite pulses are regarded as GQGs. This
shows that we can coherently interpret several composite pulses for
a specific systematic error in terms of geometric phases.
In this paper, we extend our former work 
\cite{Kondo10} in order to include general systematic errors. 
The derived 
conditions are simple enough to be understood straightforwardly and applicable 
not only to GQGs, but also to gates involving dynamical phases. 
Our new formalism is applicable straightforwardly to multi-partite systems either.
As a demonstration of our formalism, we design a new composite pulse robust 
against the most important systematic errors in NMR. The 
obtained pulse sequence can be seen as a concatenation of two composite 
pulses derived in \cite{Cummins03} and has high fidelity over a wide range
 in the error parameter space. This pulse sequence cannot be constructed
by iterative expansion \cite{Tycko85, Levitt96}.

This paper is organized as follows. In Sec.~II, we present our
formalism to design unitary gates robust against general errors. 
The robustness of the GQGs against the pulse-length error
is generalized to arbitrary non-degenerate multi-level systems
in the continuous time cyclic evolution, which results in Abelian 
geometric phases.
Further, we derive robustness conditions systematically based on our theory,
after which discrete time
formalism is introduced. In Sec.~III, the developed formalism is applied 
to construct concatenated pulse sequences, which are robust against
the most important systematic errors in a two-level system.
Sec.~IV is devoted to conclusion and discussions.

\section{Robustness Condition}

First, we define {\it robustness} of a gate.
Consider the special unitary group ${\rm SU}(n)$, 
whose dimension as a group manifold is
$N:=n^2-1$, and its Lie algebra $\mathfrak{su}(n)$.
We introduce a complete set of orthogonal Hermitian
basis (generators) of $\mathfrak{su}(n)$,
$\{\tau_\mu\,|\, \tau_\mu=\tau_\mu^\dag, \mu=1,\ldots, N\}$ 
with respect to the Hilbert-Schmidt inner product,
and a time-dependent real $N$-vector
$\lambda(t)=(\lambda_1(t),\ldots, \lambda_{N}(t))$ in the
$N$-dimensional parameter manifold ${\cal M}$. By choosing a continuous
path $\lambda(t)$ in ${\cal M}$, we define a family of time-dependent
Hamiltonians
\be
H(\lambda(t)):=\lambda_\mu(t)\tau_\mu \in  \mathfrak{su}(n),
\label{defH}
\ee
whose time-evolution operator 
\be
U_\lambda(t, 0):={\cal T}\exp\[-\im\int_{0}^t\dd sH(\lambda(s))\]
\label{q}
\ee
is an element of ${\rm SU}(n)$.
Here, ${\cal T}$ denotes the time-ordered product and we set $\hbar=1$ 
and employed the Einstein summation convention for
Greek indices. $U_\lambda(t, 0)$ reduces to the identity
operator $\openone$ at $t=0$. Note that the system is a two-level system 
when $n=2$ 
and $\tau_\mu=\sigma_\mu/2$, where $\sigma_\mu$ is the $\mu$-th component 
of the Pauli matrices.

Now, let us scale $t\in[0,1]$ and require that
the time evolution operator $U_{\lambda}(1,0)$ implements
a target gate $U$ at $t=1$; $U_{\lambda}(1,0)=U$.
We define $U_\lambda(1,0)$ is a {\it robust} gate if the condition 
\be
U_{\lambda+\delta\lambda}(1,0)=U+{\cal O}(|\delta\lambda|^2)
\label{udelta}
\ee
is satisfied for a given $\delta\lambda(t)$ with
$|\delta\lambda(t)|\ll|\lambda(t)|$ for every $t$.
To find the robustness condition, let us rewrite the LHS of
Eq.~(\ref{udelta}) in the interaction picture. Consider the dynamics
under a Hamiltonian
$H(\lambda(t)+\delta\lambda(t))=H(\lambda(t))+H(\delta\lambda(t))$,
where $H(\delta\lambda(t))$ is regarded as a perturbation. Then, we
obtain
\be
U_{\lambda+\delta\lambda}(1,0)=UU_{\delta\lambda}^I(1,0),
\label{declambda}
\ee
where
\be
U_{\delta\lambda}^I(1,0)={\cal T}\exp\[-\im\int_0^1\dd tH_I(\delta\lambda(t))\]
\ee
is defined through the interaction picture Hamiltonian
\be
H_I(\delta\lambda(t)):=U_\lambda(t,0)^\dag H(\delta\lambda(t))U_\lambda(t,0).
\label{intH}
\ee

Equation (\ref{declambda}) requires that {\it $U_\lambda(1,0)$ 
is robust against a given error $\delta\lambda(t)$ if 
$U_{\delta\lambda}^I(1,0)=\openone$.}

To proceed further, let us rewrite $U_{\delta\lambda}^I(1,0)$ in the RHS
of Eq.~(\ref{declambda}) as the Dyson series. Then, we obtain
\be
U_{\lambda+\delta\lambda}(1,0)=U-\im U\Delta W+{\cal O}(|\delta\lambda^I|^2),
\label{expansion}
\ee
where
\be
\Delta W:=\int_0^1\dd tH_I(\delta\lambda(t))
\ee
and
$\delta \lambda^I$ is defined implicitly as a solution of $H_I(\delta \lambda)=H(\delta\lambda^I)$. 
Since ${\cal O}(|\delta\lambda^{I}|^2)={\cal O}(|\delta\lambda|^2)$ from Eq.~(\ref{intH}), 
$U_\lambda(1,0)$ is robust 
against given $\delta \lambda$  to
${\cal O}(|\delta\lambda|^2)$ if and only if
\be
\Delta W=0.
\label{lt}
\ee
Note that the robustness condition (\ref{lt}) is derived without
assuming an explicit form of 
the Hamiltonian and is applicable to many physical systems.

\subsection{Classification}

So far, we have considered general $\delta \lambda(t)$. From now on, 
we restrict ourselves within systematic (deterministic)
errors for $\delta \lambda(t)$. By definition of a
systematic error, $\delta\lambda(t)$ takes a form
\be
\delta\lambda_\mu(t)=F_\mu(\lambda(t)),
\label{fl}
\ee
where $F=(F_1,\ldots, F_N)$ is an unknown vector function defined in
${\cal M}$. 
We assume the RHS of Eq.~(\ref{fl}) admits an expansion
\be
\!\!F_\mu(\lambda(t))=f_\mu+f_{\mu\nu}\lambda_\nu(t)+
f_{\mu\nu\rho}\lambda_\nu(t)\lambda_\rho(t)+\cdots,
\label{exp}
\ee
where  $f_\mu, f_{\mu\nu}, f_{\mu\nu\rho},\ldots$ are constant
tensors. Hereafter we do not write the time-dependence of functions
explicitly to simplify equations. Substitution of Eq.~(\ref{exp}) into
Eq.~(\ref{lt}) shows that Eq.~(\ref{lt}) is satisfied for any $F$ if
\be
\!\!\!
\int_0^1\dd t\tilde{\tau}_\mu=0,
\int_0^1\dd t\tilde{\tau}_\mu \lambda_\nu=0,
\int_0^1\dd t\tilde{\tau}_\mu \lambda_\nu \lambda_\rho=0,\dots
\label{cd}
\ee
hold simultaneously. Here, we utilized the expression (\ref{defH}) for
$H(\delta\lambda(t))$ and introduced generators 
$\tilde{\tau}_\mu(t)$ in the interaction picture:
\be
\tilde{\tau}_\mu(t):=U_\lambda(t,0)^\dag\tau_\mu U_\lambda(t,0).
\ee

The first condition in Eq.~(\ref{cd}) requires that the effect of
a constant error $F_{\mu}= f_{\mu}$ on the time-evolution
operator vanishes. 

We turn to the second condition in Eq.~(\ref{cd}). To find its 
implication, $f_{\mu\nu}$ is decomposed into the sum of tensors,
each of which is an irreducible representation of ${\mathfrak{su}}(n)$:
\be
f_{\mu\nu}&=&\delta_{\mu\nu}f_{\rho\rho}/N
+(f_{\mu\nu}-f_{\nu\mu})/2\nonumber\\
&+&\[(f_{\mu\nu}+f_{\nu\mu})/2-\delta_{\mu\nu}f_{\rho\rho}/N\].
\label{ir}
\ee
Employing an analogy to fluid mechanics, each irreducible tensor in the
RHS can be thought of as an error which causes a uniform expansion,
rotation and torsion of $\lambda(t)$, respectively.
According to the decomposition (\ref{ir}), we find
\begin{subequations}
\be
&& \!\!\!\! \int_0^1\dd t\tilde{\tau}_\mu\lambda_\mu
=\int_0^1\dd tU_\lambda(t,0)^\dag H(\lambda(t))U_\lambda(t,0)=0,
\label{s}\\
&&\!\!\!\! 
\int_0^1\dd t\(\tilde{\tau}_\mu\lambda_\nu-\tilde{\tau}_\nu\lambda_\mu\)=0,
\label{v}\\
&&\!\!\!\! 
\int_0^1\dd t \[\(\tilde{\tau}_\mu \lambda_\nu
+\tilde{\tau}_\nu\lambda_\mu\)/2
-\delta_{\mu\nu}\tilde{\tau}_\rho\lambda_\rho/N\]=0
\label{t}
\ee
\label{svt}
\end{subequations}
$\hspace{-5pt}$as the sufficient conditions for the robustness against
the corresponding errors. 

\subsection{Geometric Phase Gate and Norm Error Compensation}
Let us consider the case when there is only a norm error in $\lambda$
exists. In this case, Eq.~(\ref{exp}) is reduced to
\begin{eqnarray}
\!\!F_\mu(\lambda(t))= f_{\rho\rho} \lambda_\mu(t),
\end{eqnarray}
and we have to consider only 
$\int_0^1 {\rm d}t \tilde{\tau}_\mu \lambda_\mu$ in Eq.~(\ref{s})
for evaluating its robustness. 

Let us introduce $\ket{\psi_a(t)}:=U_{\lambda}(t,0)\ket{\psi_a(0)}$ 
and assume
non-degeneracy of eigenvalues throughout time-evolution. 
By taking the 
expectation value of Eq.~(\ref{s}) with respect to $\ket{\psi_a(0)}$, 
we find that the dynamical phases \cite{Aharonov87}
\be
\gamma_{\rm d}^a:=-\int_0^1\dd t\bra{\psi_a(t)}H(\lambda(t))\ket{\psi_a(t)}
\label{dpvanish}
\ee
must vanish for all $a$.
Now let us consider a case in which $\ket{\psi_a(0)}$ is a cyclic state of 
$U_\lambda(1,0)$ with a phase $\gamma^a\in\R$ \cite{Aharonov87},
that is, $\ket{\psi_a(0)}$ is an eigenvector of 
$U_\lambda(1,0)$ with the eigenvalue $\e^{\im\gamma^a}$,
\be
\ket{\psi_a(1)}=U_\lambda(1,0)\ket{\psi_a(0)}=\e^{\im\gamma^a}\ket{\psi_a(0)}.
\ee 
We have the spectral decomposition of $U_\lambda(1,0)$
in terms of these mutually orthogonal
$\{\ket{\psi_a(0)}\}_{a=1}^n$ as
\be
U_\lambda(1,0)=\sum_{a=1}^n\e^{\im\gamma^a}\ket{\psi_a(0)}\bra{\psi_a(0)}.
\ee
Let us recall that a cyclic state admits the Aharonov-Anandan phase
\cite{Aharonov87}
\be
\gamma_{\rm g}^a:=\gamma^a-\gamma_{\rm d}^a.
\ee
Thus, we realize a non-trivial unitary gate ($\gamma^a\not\equiv0, \mod 2\pi$ for some $a$) robust against the error on the norm
of the vector $\lambda(t)$, if it has a nonvanishing geometric contribution
$\gamma_{\rm g}^a$ under the condition (\ref{s}), 
which leads to $\gamma_{\rm d}^a=0$ for all $a$. 
This observation confirms that our formalism is a proper continuous time generalization of the previous work \cite{Kondo10},
which revealed that composite pulses robust against the pulse-length
error are the GQGs.

\subsection{Discretization}

Next, let us divide the temporal interval $[0,1]$ into $k$
intervals, in each of which the Hamiltonian is constant. 
With this piecewise constant Hamiltonian, 
time-ordering in the time-evolution operator $U_{\lambda}(t,0)$ 
is simply an ordered product of time-evolution operators defined for each time-independent
Hamiltonian. This means that we restrict ourselves within the gate $U$ which
is decomposed into a product
\be
U=U_{\lambda^k}(t_k, t_{k-1})\cdots U_{\lambda^{1}}(t_{1}, t_0),
\label{decU}
\ee
where $0 = t_0 < t_1 < \dots < t_{k-1}< t_k=1$ and $\lambda^i$ is 
the set of constant
parameters in the Hamiltonian $H(\lambda(t_i))$ corresponding to the interval
$[t_{i-1}, t_i]$. Then, by definition, the unitary operator
in the presence of an error $\delta\lambda(t)$ has a similar decomposition
\be
\!\!\!\!
U_{\lambda+\delta\lambda}(1,0)=U_{\lambda^k+\delta\lambda^k}(t_k,
t_{k-1})\cdots U_{\lambda^{1}+\delta\lambda^1}(t_{1}, t_0),
\label{decUdelta}
\ee
where $\delta\lambda^i$ is the deviation of the parameters in
the interval
$[t_{i-1},t_i]$. Note that $\delta \lambda^i$ is independent of
time when $\lambda^i$ itself is independent of time.
We write
$U_{\lambda^i}(t_i, t_{i-1})$ as
\be\label{ul}
U_{\lambda^i}(t_i, t_{i-1})\rightarrow R(m^i):=\exp\(-\im m_\mu^i\tau_\mu\),
\ee
where $m^i:=\lambda^i(t_i-t_{i-1})$. By denoting the corresponding
error as $\delta m^i$, the unitary operator
$U_{\lambda^i+\delta\lambda^i}(t_i, t_{i-1})$ is replaced similarly as
\be
U_{\lambda^i+\delta\lambda^i}(t_i, t_{i-1})\rightarrow R(m^i+\delta m^i).
\ee
Let us introduce
\be
W^i:=m^i_\mu\tau_\mu
\quad
{\rm and}
\quad
\delta W^i:=\delta m^i_\mu\tau_\mu,
\ee
and recall the well-known formula
\be
{\rm e}^{A+B}&=&\int_0^1\dd x\delta(1-x){\rm e}^{Ax}\nonumber\\
&+&\int_0^1\dd x\int_0^1\dd y\delta(1-x-y){\rm e}^{Ax}B{\rm e}^{Ay}+\cdots
\label{formula}
\ee
for 
matrices $A$ and $B$ of the same dimension. Then, by setting $A=-\im W^i$ and
$B=-\im\delta W^i$ and neglecting higher order terms with respect to
$B$, we obtain
\be
R(m^i+\delta m^i)\approx R(m^i)-\im R(m^i) \delta W^i_I,
\ee
where
\be
\delta W^i_I:=\int_0^1\dd x\,\e^{\im x W^i}\delta W^i\e^{-\im x W^i}.
\label{discHint}
\ee
Thus, it follows that $\Delta W$ defined in Eq.~(\ref{expansion}) is given by
\be
\Delta W=\sum_{i=1}^k{V^{i-1}}^\dag\delta W^i_IV^{i-1},
\label{discDH}
\ee
with
\be
V^i=R(m^i)\cdots R(m^1)
\quad
{\rm for}
\quad
i=1,2,\dots,
\ee
and $V^0=\openone$.
Note that we utilized the identity $R(m^k)\cdots
R(m^i)=U{V^{i-1}}^\dag$ to derive Eq.~(\ref{discDH}).
Characterization of error (14) is applied under this discretization.

In case $[W^i, \delta W^i]=0$ for all $i$, we easily find
\be
\delta m^i_\mu=\epsilon m^i_\mu,
\ee
which is the error generating the uniform expansion of the norm of the
parameter vector $m_\mu$.
 Then, we observe further simplification of $\Delta W$:
\be
\Delta W=\epsilon\sum_{i=1}^k{V^{i-1}}^\dag W^iV^{i-1}.
\label{DHcom}
\ee
This is nothing but the central quantity considered in \cite{Kondo10},
in which the usefulness of Eq.~(\ref{DHcom}) has been presented in detail. 

\section{Application: Concatenated Composite Pulse}
\label{app}
As an example of our discretization formalism, with
NMR and similar systems in mind,
we construct an ${\rm SU}(2)$ gate robust against two errors 
defined by the following error operator
\be
\hspace{-8pt}
\delta W^{i}=\epsilon W^{i}+\epsilon^\prime|m^{i}| \sigma_3/2.
\label{OR}
\ee
The first term in the RHS causes the pulse-length error while the second 
term causes the off-resonance error in the terminology adopted from NMR. 
The error operator (\ref{OR}) simultaneously
represents the most important errors inherent 
in quantum control in NMR and a system with an analogous Hamiltonian.

Bearing the situation in NMR setups in mind, where there is no $\sigma_3$ in the Hamiltonian (\ref{defH}), 
 we construct gates robust against errors (\ref{OR}) under the restriction
\be
m^{i}=\theta_{i}(\cos\phi_{i},
\sin\phi_{i},0),
\qquad
\theta_i\ge0.
\ee

Now, we concatenate two composite gates, each of which is composed of
three simple unitary operations of the form (\ref{ul}).
One is the pulse sequence called Compensation for Off-Resonance with a Pulse SEquence (CORPSE), which is robust against the off-resonance error and consists of three commutative pulses 
 \cite{Cummins03}. In other words, given a target $U=R(m)$ with  $m=\theta(\cos\phi, \sin\phi,0)$,
 the elementary pulses in CORPSE are given by
 \be
\phi_{1}=\phi_{2}+\pi=\phi_3=\phi,
\label{pp}
\ee
and
\be
&&\theta_{1}=\theta/2-\kappa + 2n_{1}\pi,\nonumber\\
&&\theta_{2}=-2\kappa + 2n_{2}\pi,\nonumber\\
&&\theta_{3}=\theta/2-\kappa + 2n_{3}\pi,
\ee
where $n_{1},n_{2},n_{3}\in\mathbb{Z}$ are chosen so that $\theta_i\ge0$ and
\be
\kappa:=\arcsin\frac{\sin(\theta/2)}{2}.
\label{CORPSE}
\ee
This pulse was 
obtained in \cite{Cummins03} by making use of quaternion algebra. 
Its robustness against
random telegraph noise is examined in \cite{Mottonen06} in comparison
with other pulse sequences. CORPSE is used to compensate for
the off-resonance error in NMR experiment \cite{Cummins00}.
 
The other composite pulse is the SCROFULOUS, which is an acronym of Short Composite ROtation For Undoing Length Over and Under Shoot. SCROFULOUS is robust against the pulse-length error and made of a $\pi$-pulse sandwiched between two identical pulses \cite{Cummins03}. More precisely, three elementary pulses satisfy
\begin{equation}
m^1=m^3,
\quad
\theta_2=\pi,
\quad
\theta_1\cos\(\phi_1-\phi_2\)+\pi/2=0.
\label{scr}
\end{equation}
SCROFULOUS is a generalization of a
composite gate proposed in \cite{Tycko85b}, which has experimental confirmation for its robustness \cite{Cummins00}.

Now we are ready to construct the concatenated composite pulse
by combining three CORPSE gates to form a SCROFULOUS gate.
First, from Eq.~(\ref{discHint}), we find
\be
\delta W_I^{i}=\epsilon W^{i}+\epsilon^\prime R(m^{i})^\dag\sigma_3\sin(\theta_{i}/2).
\label{deltaWI}
\ee
for the error operator (\ref{OR}).
For this, Eq.~(\ref{discDH}) for $k=3$ gives
\begin{eqnarray}
\Delta W&=&R(m^{1})^\dagger R(m^{2})^\dagger \delta W_I^{3} R(m^{2})R(m^{1})
\nonumber\\
& +& R(m^{1})^\dag \delta W_I^{2}R(m^{1})+ \delta W_I^{1}.
\end{eqnarray}
For notational convenience, we work with
\be
U\Delta W=\epsilon R(m^{3}) S+\epsilon^\prime T,
\ee
instead of $\Delta W$, where 
\be
\label{sp}
S&=&W^{3}R(m^{2})R(m^{1})+R(m^{2})W^{2}R(m^{1})\nonumber\\
&+&R(m^{2})R(m^{1})W^{1}
\ee
and
\be
&&T=\sin(\theta_3/2)\sigma_3R(m^2)R(m^1)+\sin(\theta_2/2)R(m^3)\sigma_3R(m^1)\nonumber\\
&&\hspace{11pt}+\sin(\theta_1/2)R(m^3)R(m^2)\sigma_3.
\label{ta}
\ee

\begin{figure*}[t]
  \begin{center}
  \begin{minipage}{1.6in}
   \includegraphics[scale=.52,clip]{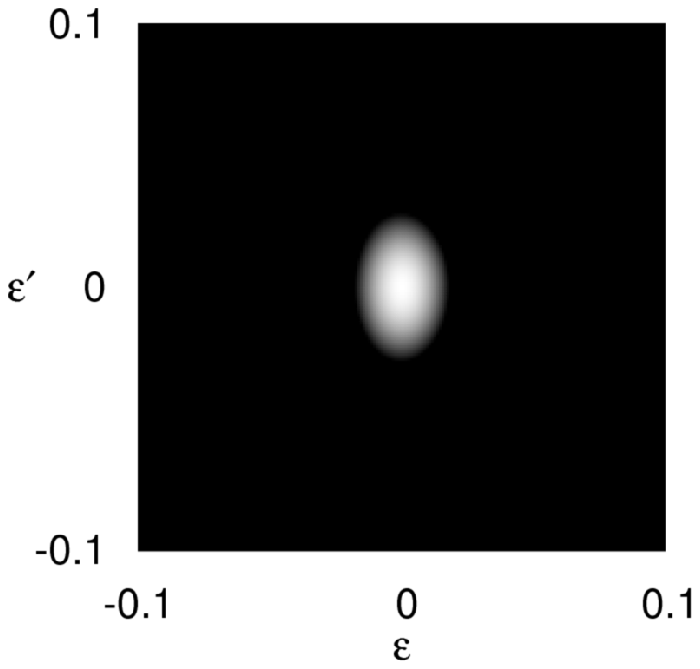}\\ 
   (a) Plain
  \end{minipage}
  \begin{minipage}{1.6in}
   \includegraphics[scale=.52,clip]{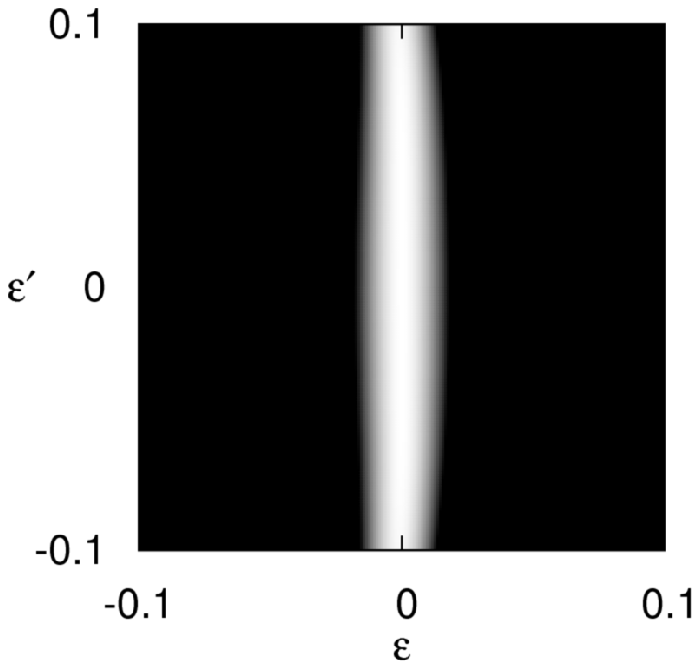}\\ 
   (b) CORPSE
  \end{minipage}
  \begin{minipage}{1.6in}
   \includegraphics[scale=.52,clip]{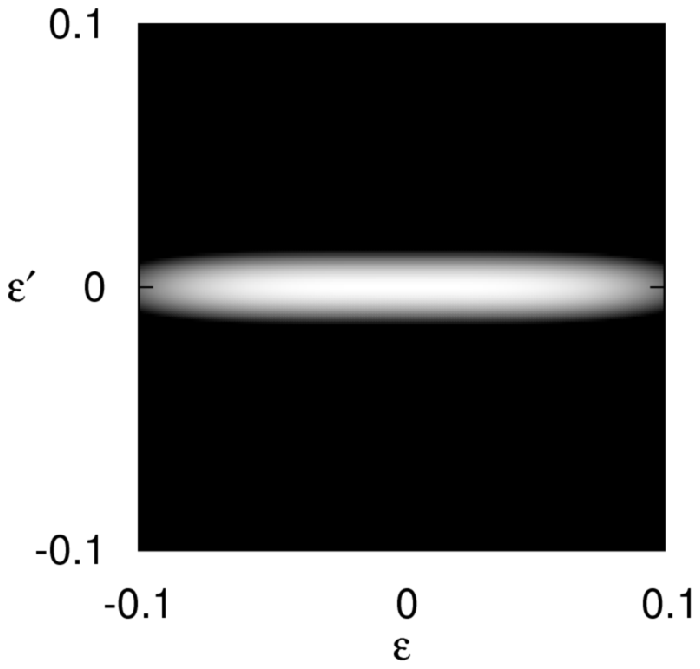}\\ 
   (c) SCROFULOUS
  \end{minipage}
\begin{minipage}{1.6in}
  \hspace{0mm}
   \includegraphics[scale=.52,clip]{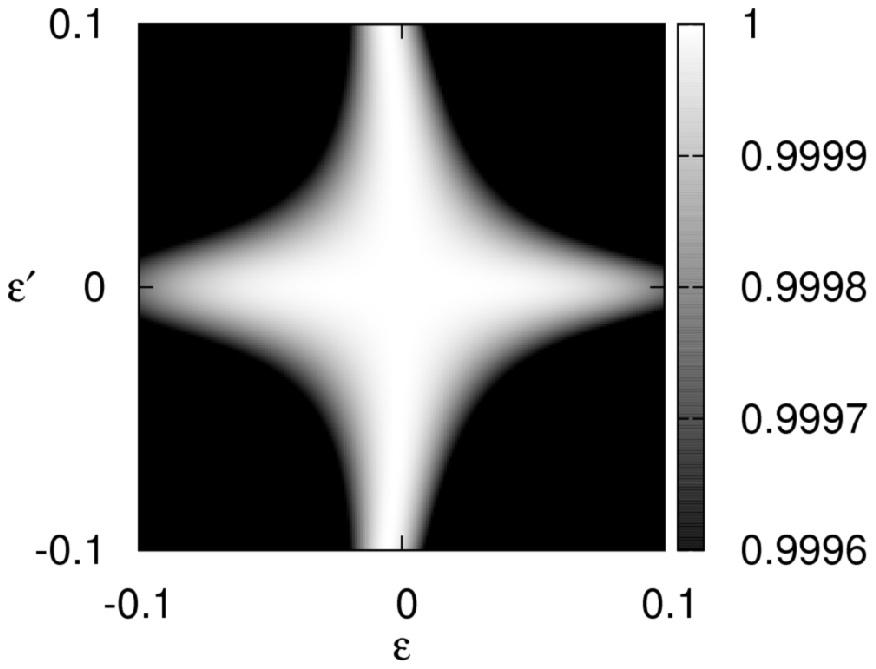}\\ 
   (d) Concatenated pulses
  \end{minipage}
\end{center}
 \vspace*{-2mm}
 \caption{Fidelity ${\cal F}$ of (a) a plain pulse
 (b) CORPSE ($n_{1} = n_{3} = 1,\ n_{2} = 2$),
 (c) SCROFULOUS and (d) the concatenated 
 pulse sequence as a function of the error strength constants $\epsilon$ for the pulse-length error and $\epsilon^\prime$ for the off-resonance error. The target unitary
 operator $U$ is $\exp(-\im\pi\sigma_2/2)$. In whiter area the fidelity ${\cal F}$ yields higher value.}
 \label{f1}
\end{figure*}

For CORPSE, we observe that $T=0$ as expected, since it is designed 
so as to compensate for the off-resonance error.
Further, from Eq.~(\ref{pp}), 
we have $[W^{i}, R(m^j)]=0$, which leads to
\be
\Delta W&=&\epsilon(W^1+W^2+W^3)\\
&=&\epsilon\[\theta/2+(n_1-n_2+n_3)\pi\](\cos\phi\sigma_1+\sin\phi\sigma_2).\nonumber
\ee
Let us choose $\{n_i\}$ so that they satisfy
\be
n_{1} - n_{2} + n_{3} = 0.
\ee
Then we have
\be
\Delta W &=&\epsilon m_\mu\tau_\mu,
\ee
which is nothing but the pulse-length error acting on the target unitary operator.
This clearly tells us that we can compensate for both systematic errors 
simultaneously,
if we use  a ConCatenated Composite Pulse (CCCP) sequence of three CORPSE 
sequences under the condition that they compose the SCROFULOUS when
combined together. We call this concatenated pulse by
CORPSE In SCROFULOUS-CCCP, or CIS-CCCP for short, in the following.  

One could alternatively try a concatenation of three 
SCROFULOUS pulses under the condition that they 
compose the CORPSE. This pulse sequence is, however, not robust in the sense 
of Eq.~(\ref{udelta}): Each constituent SCROFULOUS in the pulse 
sequence leads to $S=0$ but $T\neq\sigma_3\sin(\theta/2)$. This implies 
$\Delta W\neq \epsilon^\prime U^{\dag}\sigma_3\sin(\theta/2)$, that is, in view of the second term  in the RHS of Eq.~(\ref{deltaWI}), the error which is not compensated for by each SCROFULOUS
pulse cannot be
regarded as the off-resonance error and the overall CORPSE fails to 
eliminate it. 

We would like to emphasize that the CIS-CCCP sequence cannot be generated by 
iterative expansions \cite{Tycko85, Levitt96}.
An iterative expansion is composed of consecutive applications of 
various pulse sequences, each of which is created from a given 
pulse sequence by i) a permutation and ii) a shift of the rotation 
axes in the $xy$-plane of constituent pulses.
Then, one cannot create a CORPSE with the total rotation angle 
$\pi$ by operations i) and ii) on a generic CORPSE.
This proves impossibility of designing the CIS-CCCP by iterative expansions.

It is of interest to compare the fidelity of CIS-CCCP sequence with those of 
CORPSE and SCROFULOUS. The fidelity with respect to the target 
unitary gate $U$ is defined by the absolute value of the 
Hilbert-Schmidt inner product:
\be
{\cal F}=\frac{1}{2}\Big|\Tr\[U^\dag R(m^k+\delta m^k)\cdots R(m^1+\delta m^1)\]\Big|.
\ee
Note that $k=3$ for CORPSE and SCROFULOUS, whereas
$k=9$ for CIS-CCCP.  
 
Our interest lies in the weak error strengths region
$-0.1\le\epsilon, \epsilon^\prime\le0.1$, 
 since the accuracy threshold theorem requires the error probability less than
 ${\cal O}(10^{-3})$ for fault tolerant quantum computation \cite{Gaitan07}. 
 From Fig.~1, we immediately observe two features of the CIS-CCCP. First, the 
 CIS-CCCP has characteristics of both the SCROFULOUS and the CORPSE pulses, 
 as expected: 
The CIS-CCCP is robust along the lines $\epsilon\epsilon^\prime=0$, 
whereas the CORPSE is robust along $\epsilon=0$ and the SCROFULOUS along 
 $\epsilon^\prime=0$. Second, the whiter area, the higher fidelity region, 
of the CIS-CCCP is considerably wider than 
 those of the CORPSE and the SCROFULOUS combined together. This observation
indicates that the concatenation of composite pulses results 
in an even more robust pulse sequence.

In closing this section, let us show the difference between the CIS-CCCP 
and the composite pulse proposed by Alway and Jones~\cite{Alway07}. 
Their composite pulse also compensates the pulse-length error and the
off-resonance error simultaneously, but implements only $\pi$-pulses along an 
axis in the $xy$-plane (see Appendix \ref{a1}).
Here, we should note that any combinations of $\pi$-pulses along axes on the $xy$-plane are reduced to either $\pi$-pulse on the same plane or the pulse along
the $z$-axis:
\begin{eqnarray}
 R(m^k)\cdots R(m^1) = 
 \begin{cases}
R(m),      & k=1, 3,\dots, \\
R(m^z),      & k=2,4,\dots.
\end{cases}
\label{AJprod}
\end{eqnarray}
Here, $m^j = \pi(\cos\phi_j,\sin\phi_j,0)$ and
$m = \pi(\cos\phi,\sin\phi,0)$, where $\phi$ is read as a function of $\{\phi_j\}$.
We introduced $m^z = (0,0,2\Theta)$
with $\Theta = \sum_{j=1}^{k/2} (\phi_{2j} - \phi_{2j-1}+\pi)$.
Equation (\ref{AJprod}) can be derived by mathematical induction with respect to $k$.
Therefore, it is impossible to implement arbitrary one-qubit rotations as combinations of their composite pulse sequences. This clearly shows that we cannot realize universal gate set by using their composite pulse sequence.
In contrast, the CIS-CCCP does not have such a restriction as Eq.~(\ref{AJprod}), and implements any one-qubit unitary operation
by using Euler angles.

\section{Conclusion and Discussions}

In this paper, we proposed a simple formalism to design unitary
gates robust against systematic errors whose magnitude are unknown. 
By using this method, we systematically derived various criteria 
which admitted lucid interpretations. 

We designed a pulse sequence robust against two types of systematic errors
(\ref{OR}) simultaneously. 
We design a new composite pulse that is the 
SCROFULOUS out of the CORPSEs in order to take 
advantages of these composite pulses. Our approach 
is straightforward than the quaternion algebra that relies on brute force calculation. Our pulse sequence has controllable free parameters; cf. the 
pulse sequence proposed in \cite{Alway07}, which is also robust against 
the errors (\ref{OR}), but implements only $\pi$-pulse gate on the $xy$-plane. In contrast,
our pulse sequence realizes arbitrary one-qubit unitary gate robust 
against combined errors (\ref{OR}) and will find an important application 
in implementation of universal gate set out of low quality gates.
These features show the usefulness of our formalism as a guiding principle
to construct unitary gates robust under coexisting deterministic errors.

We would like to stress that the our scheme is applicable not only to 
NMR, but also to other physical systems since the condition
(\ref{lt}) is formulated independently of the Hamiltonian.
For example, the effective Hamiltonian of a quantronium
superconducting qubit takes the NMR form \cite{Collin04}
\begin{equation}
H = -h \nu_{R0} (\cos \xi \sigma_1 + \sin \xi \sigma_2)
\end{equation}
with an additional off-resonance error term
$h \Delta \nu \sigma_3$, where $h$ is a constant
and $\Delta \nu$ is the detuning. It was demonstrated
in \cite{Collin04} that CORPSE indeed suppresses the
off-resonance error.
A similar demonstration of the effectiveness of CORPSE
has been made for a neutral atom qubit to suppress
effective microwave detuning across the qubit ensemble \cite{Rakreungdet09}.
Our concatenated composite pulse is applicable to these systems.

Our method is also useful for designing a robust two-qubit gate. 
Nonetheless, it requires intensive analytical as well as numerical analysis 
and is beyond the scope of the present paper. 
Our preliminary result shows that a two-qubit gate with error in the
coupling strength 
between qubits may be made robust against the error by decomposing the gate 
into a relatively small number of gates, which will be reported elsewhere.

\begin{acknowledgements}
The authors wish to thank the referee for valuable comments and suggestions
which led to a significant improvement of the manuscript.
This work is supported by \lq Open Research Center\rq~Project for
 Private Universities; matching fund subsidy, MEXT, Japan. YK and 
 MN would like to thank partial supports of Grants-in-Aid for Scientific 
Research from the JSPS (Grant No.~23540470).
\end{acknowledgements}

\appendix

\section{Alway and Jones' gate\label{a1}}

In Sec.~\ref{app}, we discussed the composite $\pi$-pulse designed 
by Alway and Jones~\cite{Alway07},
which compensates both the pulse-length error and the off-resonance error.
Suppose we are to implement a pulse sequence whose rotation axis is in
the $xy$-plane.
Then the pulse sequence employed in \cite{Alway07} is robust under
the error (\ref{OR}) if and only if the rotation angle is $\pi$. 

The ``if'' part is proved by Alway and Jones~\cite{Alway07} as mentioned
above. Let us briefly reproduce their result to establish notation
and conventions.
For notational convenience, we introduce $U(\theta):=\exp\(-\im\theta\sigma_1/2\)$. (We consider arbitrary rotation axes at the end of Appendix.)
Given the target $U(\pi)$, we can design the following pulse sequence:
\be
 U_{\rm or} U_{\rm pl} U(\pi),
 \label{aj}
\ee
where we introduced two partial sequences
\be
U_{\rm pl} = \prod_{i=1}^3R(m^i),
\qquad
U_{\rm or} = \prod_{i=4}^7R(m^i),
\ee
whose elementary pulses are given as
\be
&& \theta_2 = 2\pi,\quad \theta_i = \pi\quad (i\neq 2)\nonumber\\
&& \phi_1 = \phi_3 = \phi_{\rm pl},\quad \phi_2 = 3\phi_{\rm pl},\nonumber\\
&& \phi_4 = 2\pi-\phi_6 = \pi - \phi_{\rm or},
\quad \phi_5 = -\phi_7 = -\phi_{\rm or},
\ee
with $\phi_{\rm pl} = \phi_{\rm or} = \arccos(-1/4)$.
$U_{\rm pl}U(\pi)$ is a composite pulse known as BB1~\cite{Wimperis}, which is
robust against the pulse-length error, while $U_{\rm or}U(\pi)$ is a composite pulse robust against the off-resonance error.
Note that
\be
U_{\rm pl}=U_{\rm or}=\openone
\ee
in the absence of errors.
This is the sequence proposed in \cite{Alway07}.

Now let us prove the ``only if'' part. Suppose we want to implement
$U(\theta)$, which is robust against simultaneous errors
by employing the sequence (\ref{aj}). Here we assume
$\theta$ is not fixed to $\pi$ and $\phi_{\rm pl}$ and $\phi_{\rm or}$
are adjusted so as to make the gate robust for a given $\theta$.
After simple calculation, we obtain the zeroth and the 
first order error terms as
\be
 U(\theta) - \im\epsilon\frac{\theta}{2}\sigma_1 U(\theta)
 -\im\epsilon^\prime\sin\frac{\theta}{2}\sigma_3,
\label{eq:uth}
\ee
for the target $U(\theta)$.
In the same way, we have
\be
U_{\rm pl}= \openone -2\pi\im\epsilon\cos\phi_{\rm pl}\sigma_1
\label{eq:upl}
\ee
and
\be
U_{\rm or}= \openone +4\im \epsilon^\prime\cos\phi_{\rm or}\sigma_2.
\label{eq:uor}
\ee
Taking product of (\ref{eq:uth}), (\ref{eq:upl}) and (\ref{eq:uor}) and 
evaluating the coefficients of $\epsilon$ and $\epsilon^\prime$, 
we find that this composite pulse is robust against the simultaneous
errors only if $\theta$, $\phi_{\rm pl}$ and $\phi_{\rm or}$ satisfy the 
conditions:
\begin{equation}
  2\pi\cos\phi_{\rm pl}
 + \frac{\theta}{2} = 0,
 \quad
 4\cos\phi_{\rm or}\sigma_2 U(\theta)
 - \sin\frac{\theta}{2}\sigma_3 = 0.
\end{equation}
These conditions have a unique nontrivial solution 
\be
 \phi_{\rm pl} = \phi_{\rm or} = \arccos(-1/4),
 \quad \theta = \pi,
\ee
which shows that the rotation angle of the target pulse must be $\pi$.

Clearly the rotation axis can be any direction in the $xy$-plane by
a simple redefinition of the coordinate axes or by applying the similarity
transformations around the $z$-axis to each constituent pulses. 
This completes the proof of the ``only if'' part.


\end{document}